# Lesion Segmentation in Whole-Body Multi-Tracer PET-CT Images; a Contribution to AutoPET 2024 Challenge

Mehdi Astaraki[1,2] and Simone Bendazzoli[3,4]

[1] Department of Medical Radiation Physics, Stockholm University, Stockholm, Sweden
[2] Department of Oncology-Pathology, Karolinska Institutet, Solna, Sweden
[3] Department of Biomedical Engineering and Health Systems, KTH, Huddinge, Sweden
[4] Department of CLINTEC, Karolinska Institutet, Stockholm, Sweden
`mehdi.astaraki@ki.se`

**Abstract.** The automatic segmentation of pathological regions within whole-body PET-CT volumes has the potential to streamline various clinical applications such as diagnosis, prognosis, and treatment planning. This study aims to address this challenge by contributing to the AutoPET MICCAI 2024 challenge through a proposed workflow that incorporates image preprocessing, tracer classification, and lesion segmentation steps. The implementation of this pipeline led to a significant enhancement in the segmentation accuracy of the models. This improvement is evidenced by an average overall Dice score of 0.548 across 1611 training subjects, 0.631 and 0.559 for classified FDG and PSMA subjects of the training set, and 0.792 on the preliminary testing phase dataset.

## 1 Introduction

18F-fluorodeoxyglucose positron-emission tomography (FDG-PET) is a critical imaging modality in oncology, utilizing radiolabeled glucose to visualize and quantify metabolically active tumors. The degree of FDG uptake varies across cancer types, with some tumors like prostate cancer exhibiting low uptake and others like non-small cell lung cancer demonstrating high uptake [1]. While FDG-PET effectively captures metabolic activity, the poor resolution of PET images can hinder precise tumor delineation. Additionally, normal structures such as cardiac muscles, the brain, liver, and areas of inflammation may also exhibit increased FDG uptake. Furthermore, some primary tumors with clear boundaries may show minimal FDG uptake. Consequently, PET images must be interpreted in conjunction with structural CT modality. However, the manual detection and segmentation of tumors in PET-CT is time-consuming and labor-intensive, leading to high costs and effort. This, in turn, prevents the widespread clinical adoption of quantitative image analysis beyond research settings [2].

Deep learning (DL) techniques have significantly advanced automatic segmentation of both anatomical structures and pathological regions in structural imaging modalities like CT and MRI over the past decade [3], [4], [5]. The AutoPET challenge [6], introduced at MICCAI in 2022 and reiterated in 2023, focused on evaluating the performance of automated lesion segmentation methods in whole-body FDG-PET-CT scans. The challenge organizers provided a large-scale labeled dataset of whole-body PET-CT scans from patients diagnosed with lung cancer, melanoma, or lymphoma.

This paper details our contribution to the MICCAI AutoPET 2024 challenge, where we conducted a series of experiments for automatic lesion segmentation in whole-body scans.

## 2  Materials and Methods

Following the experiences of autoPET challenges I and II, the autoPET III was established to address the crucial requirement for models to generalize effectively across various tracers and clinical settings. To facilitate this, the challenge organizer granted access to a more comprehensive PET/CT dataset comprising images acquired using two distinct tracers - Prostate-Specific Membrane Antigen (PSMA) and FDG from two separate clinical sites.

### 2.1  Studied Dataset and Task

The utilized training dataset comprises 1611 multi-institutional co-registered PET-CT volumes, with 1014 subjects from the FDG cohort and 597 from the PSMA cohort. In the FDG cohort, 513 subjects served as negative controls, while the remaining 501 had histologically confirmed diagnoses of malignant melanoma, lymphoma, or lung cancer. The PSMA cohort included pre-and/or post-therapeutic PET-CT volumes from 537 male subjects diagnosed with prostate carcinoma and 60 subjects without PSMA-avid tumor lesions. The pathologies were annotated by, first, identifying the tracer-avid tumor lesions through visual examination along with referring to clinical examination reports. The pathologies were then manually annotated in axial slices. Accordingly, the primary challenge objective was accurate binary segmentation of FDG- and PSMA-avid tumor lesions in whole-body PET/CT images.

Following the development phase, participants could evaluate their models on a limited subset of 5 cases, before submitting a final version for assessment on the 200-case test set. Participants were permitted to develop either a single model for simultaneous multi-tracer analysis or separate models for FDG and PSMA data, contingent upon accurate tracer-type classification.

### 2.2  Methods

Preprocessing:
Data preparation and preprocessing were conducted in two sequential stages prior to model training. Initially, a cropping procedure was implemented on the CT volumes to maximally preserve anatomical structures while minimizing background inclusion in training patches. This process involved a set of thresholding and connected component analysis to segment the volumes into foreground (body) and background regions, using the body skin as the delineating boundary. Subsequently, a bounding box was generated to encompass the widest extent of the segmented CT volume. This bounding box was then applied to the corresponding PET volume to ensure spatial alignment.

The second preprocessing stage focused on intensity standardization. Due to the considerable variability observed in the dynamic range of intensity values across CT volumes, a normalization scheme was employed to clip the CT intensity values within the predefined range of -800 to 800. This was followed by channel-wise Z-score normalization during the training iterations.

Multitracer Segmentation

Two models were employed for segmenting the lesions in mixed FDG-PSMA PET-CT volumes.

### SegResNet Model

The utilized MONAI SegResNet [7] architecture incorporates ResNet blocks within its encoder, each comprising two convolutional layers with normalization and skip connections. The decoder, in contrast, employs a single block per spatial level. Each decoder block initiates its operation by reducing the depth of feature maps, doubling the spatial dimension, and concatenating corresponding resolution outputs from the encoder. It's noteworthy that the implemented model excluded the Variational Autoencoder branch, adopting instead a deep supervision module.

Model training spanned 600 epochs, utilizing the following hyperparameters: variable training and validation iterations per epoch to encompass the entire dataset, a batch size of 4, an initial learning rate of 1e-4, LeakyReLU as the activation function, instance normalization, 4 levels of deep supervision, and a patch size of ($128 \times 128 \times 96$). The encoder itself consisted of six stages, characterized by (1, 3, 4, 4, 6, 6) blocks, respectively.

### nnU-Net ResENCL Model

The second segmentation pipeline utilized the nnU-Net model V2, specifically a ResNet-enhanced nnU-Net configuration (nnU-Net ResENCL) [8]. This model was employed for binary segmentation, taking two-channel CT-PET images as input. The training was conducted for 1500 epochs across five folds, with 250 training iterations and 50 validation iterations per epoch. An initial learning rate of 1e-2 and a batch size of 2 with a patch size of ($224 \times 160 \times 192$) were used, along with deep supervision. The network architecture comprised six stages in the encoder, with (1, 3, 4, 5, 6, 6, 6) blocks in each stage, and (32, 64, 128, 256, 320, 320) feature layers per stage, respectively.

Lesion Tracer Segmentation

In this second approach, we propose an alternative pipeline for segmenting cancer lesions in PET-CT volumes. This involves first identifying the radiotracer used during PET acquisition (either FDG or PSMA), and then training two separate segmentation networks, one for each PET radiotracer. The underlying hypothesis is that by eliminating intra-domain variability in the PET images from the training data, the segmentation network can better concentrate on the task of lesion segmentation.

For the radiotracer classification task, we first preprocessed the PET images by resizing them to 400×400×326 and applying Z-score normalization to the SUV-corrected values.

We employed a DenseNet-121 network for classification with a 0.2 dropout rate. The network was trained for 50 epochs, with 250 iterations per epoch, using the Adam optimizer to minimize the Binary Cross-Entropy loss function. The learning rate was set to 1e-4.

Following the initial PET radiotracer classification, we trained two separate segmentation models by dividing the training set based on the two radiotracer classes (FDG and PSMA). We conducted the experiments using the same configuration as described earlier for the segmentation task.

## 3      Experiments and Results

The predictive performance of each model was assessed by comparing the predicted segmentation masks on the validation set against the ground truth labels. For better readability, only the Dice similarity coefficient is reported in the following tables. Table 1 presents the Dice scores obtained when using the original dataset, while Table 2 showcases the performance of the same models on the preprocessed dataset. It's important to note that both models were trained using the same cross-validation splits.

**Table 1.** Segmentation performance over the original dataset

| Model | Dice Metric ($\mu \pm \sigma$) | | | | | |
|---|---|---|---|---|---|---|
|  | Fold0 | Fold1 | Fold2 | Fold3 | Fold3 | Mean |
| SegRes-Net | 0.442±0.429 | 0.481±0.382 | 0.442±0.369 | 0.459±0.417 | 0.461±0.373 | 0.457±0.388 |
| ResENCL | 0.470±0.367 | 0.520±0.369 | 0.491±0.370 | 0.473±0.382 | 0.506±0.364 | 0.492±0.369 |

**Table 2.** Segmentation performance over the preprocessed dataset

| Model | Dice Metric ($\mu \pm \sigma$) | | | | | |
|---|---|---|---|---|---|---|
|  | Fold0 | Fold1 | Fold2 | Fold3 | Fold3 | Mean |
| SegRes-Net | 0.489±0.352 | 0.531±0.376 | 0.524±0.365 | 0.469±0.395 | 0.489±0.403 | 0.500±0.382 |
| ResENCL | 0.545±0.345 | 0.599±0.328 | 0.545±0.347 | 0.514±0.374 | 0.542±+0.349 | 0.548±0.350 |

The results in Tables 1 and 2 demonstrate that the proposed preprocessing steps significantly enhanced segmentation accuracy. Furthermore, the nnU-Net ResENCL model consistently outperformed the SegResNet model. Accordingly, the configuration yielding superior performance was selected as the final Multitracer pipeline for evaluation on the preliminary test set, resulting in an average Dice metric of 0.792. Figure 1 provides visual examples of the ResENCL model's performance on the preprocessed dataset.

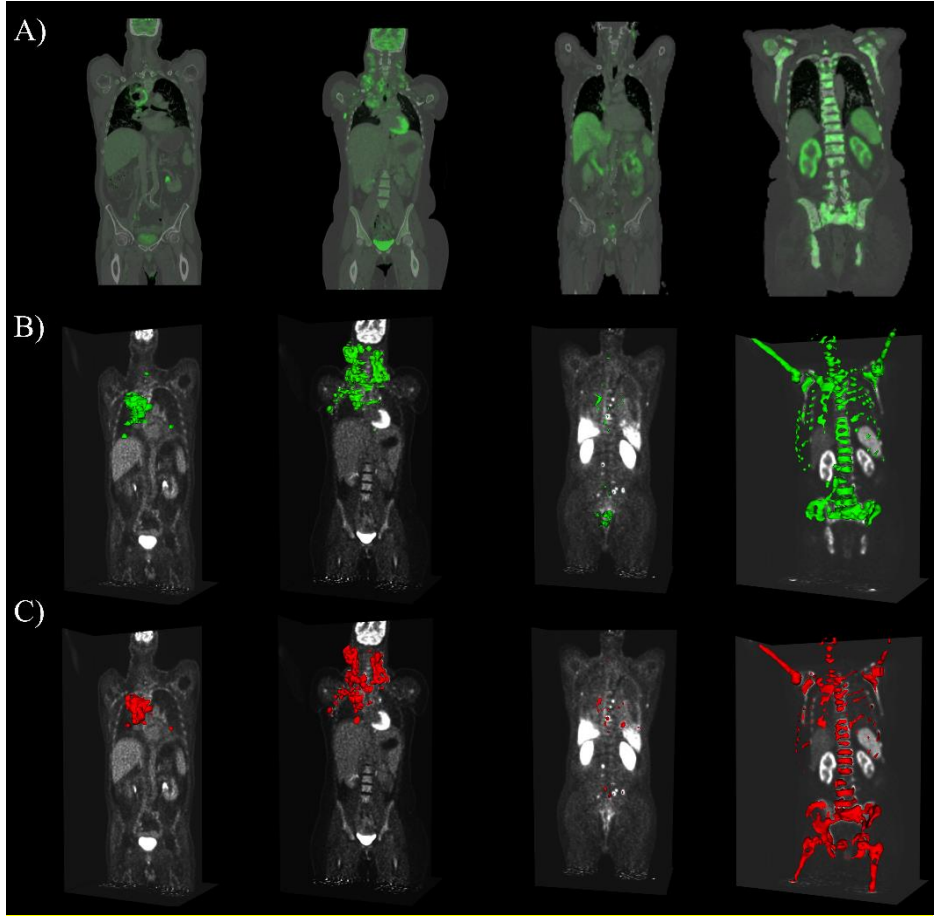

**Fig. 1.** Illustrative examples of segmented tumors across the validation sets, showcasing both FDG and PSMA cases. (A) Fused PET-CT images. (B) Ground truth tumor delineations. (C) Predicted tumor masks generated by the ResENCL model on the preprocessed dataset. The first two columns present FDG subjects, while the latter two columns display PSMA cases.

**Table 3.** Classification performance

| Model | ROC-AUC | | | | | |
|---|---|---|---|---|---|---|
| | Fold0 | Fold1 | Fold2 | Fold3 | Fold4 | Mean |
| DenseNet-121 | 0.987 | 0.995 | 0.974 | 0.991 | 0.984 | 0.986 |

Table 3 presents the average ROC-AUC score for each of the five folds in the classification task, conducted as part of the *Lesion Tracer Segmentation* approach. After completing the initial classification step, we proceeded with tracer-aware segmentation. The results from the two distinct segmentation models (evaluated only on fold 0 and fold 1) are summarized in Table 4.

**Table 4.** Segmentation performance of independent models for FDG and PSMA tracers

| Model | Dice | | |
|---|---|---|---|
| | Fold0 | Fold1 | Mean |
| FDG nnUNet-ResEnc | 0.608 | 0.608 | 0.631 |
| PSMA nnUNet-ResEnc | 0.516 | 0.601 | 0.559 |

## 4   Discussion

The AutoPET2024 challenge sought to assess the generalization capabilities of automated lesion segmentation models in whole-body PET-CT volumes by combining datasets from two major PET tracers: FDG and PSMA. This study investigated the impact of a preprocessing workflow on the delineation accuracy of established segmentation models, namely SegResNet and nnU-Net. Specific preprocessing steps examined included maximal cropping of CT volumes around anatomical structures, CT intensity normalization, and PET tracer classification.

Our experiments consistently demonstrated the superior performance of the proposed pipeline compared to the baseline segmentation models using raw data. Quantitative metrics revealed an average improvement of 4.5 percent in terms of the Dice metric. Additionally, the analyses highlighted the inferior performance of SegResNet relative to nnU-Net, possibly attributable to the experimental configuration of model architecture and training protocols. However, integrating SegResNet into self-configuration pipelines such as Auto3DSeg could potentially lead to optimization of model parameters and training strategies.

Visual inspection of the predicted masks indicated that the model was capable of localizing pathological regions, although instances of over- and under-segmentation were observed. This issue may potentially be mitigated by applying a set of rule-based criteria as a post-processing step. More importantly, the presence of healthy structures exhibiting abnormal FDG and PSMA uptake led to a high number of false positive detections. A potential strategy to reduce this undesirable effect is to incorporate anatomical structures as prior information into the segmentation pipeline. However, a major obstacle to implementing this step was the inference time, which was constrained to five minutes per subject on slow computational resources. Addressing these aforementioned issues is, therefore, a priority for our future work.

When analyzing the *Lesion Tracer Segmentation* approach, we observe significant improvements in segmentation performance by training two separate models based on the radiotracer type. The high robustness of the prior classification model (~0.99 AUC) highlights the potential benefits of this approach. However, due to time constraints during testing, we were unable to thoroughly evaluate this method in the context of the challenge. Accordingly, the *Multitracer Segmentation* model trained on combined FDG and PSMA tracers was employed for the final testing phase evaluation.

**Acknowledgments.** This study was supported by the Cancer Research Funds of Radiumhemmet. We gratefully acknowledge the Swedish Cancer Society and The Swedish Research Council (2020-04618). Our appreciation also goes to Stockholm Medical Image Laboratory and Education (SMILE) for providing access to their computational resources.

**Disclosure of Interests.** The authors have no competing interests to declare that are relevant to the content of this article.

# References

[1]  K. Hirata and N. Tamaki, "Quantitative FDG PET Assessment for Oncology Therapy," *Cancers (Basel)*, vol. 13, no. 4, pp. 1–12, Feb. 2021, doi: 10.3390/CANCERS13040869.

[2]  N. Aide, C. Lasnon, C. Desmonts, I. S. Armstrong, M. D. Walker, and D. R. McGowan, "Advances in PET/CT Technology: An Update," *Semin Nucl Med*, vol. 52, no. 3, pp. 286–301, May 2022, doi: 10.1053/J.SEMNUCLMED.2021.10.005.

[3]  M. Antonelli *et al.*, "The Medical Segmentation Decathlon," Jun. 2021, doi: https://doi.org/10.48550/arXiv.2106.05735.

[4]  F. Isensee, P. F. Jaeger, S. A. A. Kohl, J. Petersen, and K. H. Maier-Hein, "nnU-Net: a self-configuring method for deep learning-based biomedical image segmentation," *Nature Methods 2020 18:2*, vol. 18, no. 2, pp. 203–211, Dec. 2020, doi: https://doi.org/10.1038/s41592-020-01008-z.

[5]  J. Wasserthal *et al.*, "TotalSegmentator: Robust Segmentation of 104 Anatomic Structures in CT Images," *Radiol Artif Intell*, vol. 5, no. 5, Jul. 2023.

[6]  S. Gatidis *et al.*, "A whole-body FDG-PET/CT Dataset with manually annotated Tumor Lesions," *Scientific Data 2022 9:1*, vol. 9, no. 1, pp. 1–7, Oct. 2022, doi: 10.1038/s41597-022-01718-3.

[7]  A. Myronenko, "3D MRI Brain Tumor Segmentation Using Autoencoder Regularization," Springer, Cham, 2019, pp. 311–320. doi: 10.1007/978-3-030-11726-9_28.

[8]  F. Isensee *et al.*, "nnU-Net Revisited: A Call for Rigorous Validation in 3D Medical Image Segmentation," Apr. 2024, Accessed: Jul. 29, 2024. [Online]. Available: https://arxiv.org/abs/2404.09556v2